# Low energy cluster beam deposited BN films as the cascade for Field Emission

*Song Fengqi¯, Zhang Lu, Zhu Lianzhong, Ge Jun, Wang Guanghou**

National laboratory of Solid State Microstructures, Department of Physics, Nanjing University, Nanjing, 210093 P. R. China

**Abstract:**

The atomic deposited BN films with the thickness of nanometers (ABN) were prepared by radio frequency magnetron sputtering method and the nanostructured BN films (CBN) were prepared by Low Energy Cluster Beam Deposition. UV-Vis Absorption measurement proves the band gap of 4.27eV and field emission of the BN films were carried out. F-N plots of all the samples give a good fitting and demonstrate the F-N tunneling of the emission process. The emission of ABN begins at the electric field of 14.6 V/μm while that of CBN starts at 5.10V/μm. Emission current density of 1mA/cm$^2$ for ABN needs the field of 20V/μm while that of CBN needs only 12.1V/μm. The cluster-deposited BN on n-type Silicon substrate proves a good performance in terms of the lower gauge voltage, more emission sites and higher electron intensity and seems a promising substitute for the cascade of Field Emission.

¯ Email: Jackkiesong@vip.sina.com
Telephone: +86-25-83595082 +86-25-83685076, Fax: +86-25-83595535

## Introduction

A considerable effort has been devoted to the development of the cold cascade as a key device for field emission flat panel displays. It is the higher emission current and lower gauge voltage that stands for better performance in field emission. Since the 1930s much attention has been given to the emission properties of metal and the Spindt-type field emitter has been in mass production for many years [1]. The negative electron affinity (NEA) surface of diamond was discovered [2] when the research on the field emission of diamond-like material began. Boron nitride is a wide band semiconductor with its excellent performance in chemical stability, hardness and thermal conductivity, as well as NEA, which provide potential application in field electron emission. The BN coating processing has been accepted as a practical tool to improve the material performance. Moreover, a uniform film of BN is relatively easier to obtain and many methods have been engaged in producing BN thin films, such as PACVD [3, 4], RF-sputtering [5, 6] and Laser Ablation [7] etc. The band structure and the field emission properties have been characterized after the successful preparation of thickness-controlled and structure-controlled BN nanofilms [4, 6]. The recent experiments by STM have related to the emission performance with the detailed structure on nanoscale [8]. The Fowler-Nordheim equation is shown below,

$$I(F-N) = \alpha \bullet \frac{1.54 \times 10^{-6} \cdot E^2}{\Phi} \bullet \exp(-6.87 \times 10^7 \cdot \Phi^{3/2} v(y)/E) \qquad (1)$$

Where $\alpha$ is the emission area in cm$^2$, $\phi$ is the work function in eV, $v$ and $y$ are expressed below

$$y = 3.79 \times 10^{-4} E^{0.5} / \Phi \qquad (2)$$

and $\quad v = 0.95 - y^2 \qquad (3)$

The field can be written as

$$E = \beta V / d \qquad (4)$$

where β stands for field enhancement factor and d is the distance between the cascade and the collector. It is known that the work function ϕ, the field enhancement factor β and the emission area α determine the final emission properties of the cascade material. In order to improve the emission performance of a cascade material, it is necessary to lower ϕ and enlarger β and α. A large band gap and NEA provide the possibility to reduce the surface potential barrier; larger surface roughness gives larger β, while larger α needs more emission sites. In this report we present structural and field emission studies of the nanostructured films prepared by Low Energy Cluster Beam Deposition (LECBD) in comparison with those of the BN films by RF-sputtering.

**Experimental**

The CBNs were prepared in our home-built Ultra High Vacuum Cluster Beam System (UHV-CBS), which could be divided into three parts i.e. the cluster source, the sample chamber and the reflective TOF mass spectrometer. The B and N atoms were sputtered from the target by the radio frequency magnetron method and they were aggregated to form the clusters after collisions with the buffer gas of the Ar and He mixture. By differentially pumping the beam was injected to the next chamber. The beam composition could be monitored by the mass spectrum and the cluster beam was deposited on the substrates in the sample chamber. Ion beam etching and in-situ annealing could also be carried out in the system. The base pressure of the sample chamber was $2\times10^{-7}$ Torr. A RF-sputtering target was also set up in the sample chamber and was used to produce ABN.

The field emission properties were measured by the configuration shown in Fig 1. The sample was stick to the sample holder by a carbon disk and the distance between the sample and the electron collector could be changed by the micrometer screw with a precision of 20 micrometers. The base pressure of this system was $1.0\times10^{-6}$ Torr. The emission current was collected and measured by the mA-meter and nA-meter.

Both CBN and ABN samples were prepared on n-type Silicon chips and also on silica glasses for UV-Vis absorption measurements. Fig 2 shows the UV-Vis Absorption Spectrum of CBN. The bandwidth is 4.27eV, similar to that of ABN. SAED and XRD result shows the non-crystalline structure of the prepared films. Fig 3a and b show the SEM images of ABN and CBN respectively. In the former only very few blow-ups can be seen while many fluctuations are observed in the SEM images of CBN where the spots are about 100nm.

The field emission was measured and the results are shown in Fig 4. Fig 4a and Fig 4b are the *log (I-V)* curves of ABN and CBN respectively. The emission currents increase from the order of nano-Ampere at low voltage and reach milli-Ampere at higher voltage. The ABN curve increases sharply while the current for the CBN grows more steadily. The saturated current density of ABN is much lower than the CBN. The *ln(I/(V\*V))* are calculated and plotted to 1000/V (i.e. F-N Plot as shown in Fig 4c Fig 4d) and straight lines appear in the low electric-field region, indicating the existence of F-N tunneling.

**Discussions**

From Fig 4c and Fig 4d, we can find that the F-N tunneling for ABN occurs at 14.4V/μm, while that of CBN begin at 7.14V/μm and reach 1mA at 12.2V/μm (the sample area is about 0.5cm²). The field enhancement factor and the emission area are calculated from F-N plot. $\beta = 64.3$ and $\alpha = 3.30 \times 10^{-10} cm^2$ for ABN, whereas $\beta = 381$ and $\alpha = 9.6 \times 10^{-8} cm^2$ for the CBN. These measurements demonstrate that cluster-deposited BN films has an excellent emission properties .

In comparison with atomic-deposition films, cluster films by LECBD have several merits (1) Nano-clusters provide larger band width, which will elevate the energy of the injected electrons from the substrate and benefit tunneling over the surface barrier easily; (2) LECBD gives more emission sites. In the experiments the emission area $\alpha = 9.6 \times 10^{-8} cm^2$ of the CBN is two orders higher than that of ABN.

**Conclusion**

BN films are prepared by RF-sputtering and LECBD methods and their field emissions were measured. The cluster-deposited BN films show better FE in terms of lower gauge voltage, more emission sites and higher current intensity. It is possible to select the clusters with certain NEA because the cluster sizes could be selected and then nanoclusters of different sizes will have different structures with different electronic band structures. The cluster-deposited BN on n-type Silicon substrate seems a promising substitute for the cascade of Field Emission.

**Acknowledgement**

This work was financially supported by the National Natural Science Foundation of China (Grant No. 90206033, 10274031, 10021001, 10474030, 60478012), the Foundation for University

Key Teacher by the Ministry of Education of China (Grant No. GG-430-10284-1043), as well as the Analysis and Measurement Foundation of Nanjing University. Our thanks were also extended to Huang Haibo of Southeast University for his help in SEM observation.

**References**

1. C. A. Spindt et al J.Appl.Phys. **47**, no 12 (1976)

2. M.J.Powers et al Appl. Phys. Lett. **67**,3912 (1995)

3. Chiharu Kihuma et al J. Vac. Sci. Technol. B **19** 3 (2001)

4. Y. Yokota S. Tagawa S. T. Sugino Applied Surface Science **146** 193 (1999)

5. G. R. Gu et al. Chin. Phys. Lett **20** 947 2003

6. X. Z. Ding, X.T. Zeng H. Xie Thin Solid Films **429** 22 2003

7. H. H. Busta, R.W. Pryor J. Appl. Phys. **82** 10 (1997)

8. T. Sugino, C. Kimura, T. Yamamoto, S. Funakawa Diamond and Related Materials **12** 464 2003

Figures' Caption

1. The instrument used in FE measurement. The sample is stick to the sample bed by a carbon disk and the distance between the sample and the electron collector can be changed by the micrometer screw with a precision of 20 micrometers. The base pressure of this system is *1.0×10$^{-6}$* Torr. The emission current is collected and measured by the mA-meter and nA-meter, which are protected by resistances of 1KΩ and 1000KΩ respectively.

2. The UV-Vis Absorption Spectrum of CBN. Calculating by Tauc's plot, the bandwidth is 4.27eV.

3.       SEM images of ABN (a) and CBN (b) respectively.

4.       A and B are the log I-V curves of ABN and CBN respectively. Please note that in B the current is given by $\mu A$. C and D is F-N plot of ABN and CBN respectively. α and β can be calculated and they are given in the body of the paper.

Fig 1

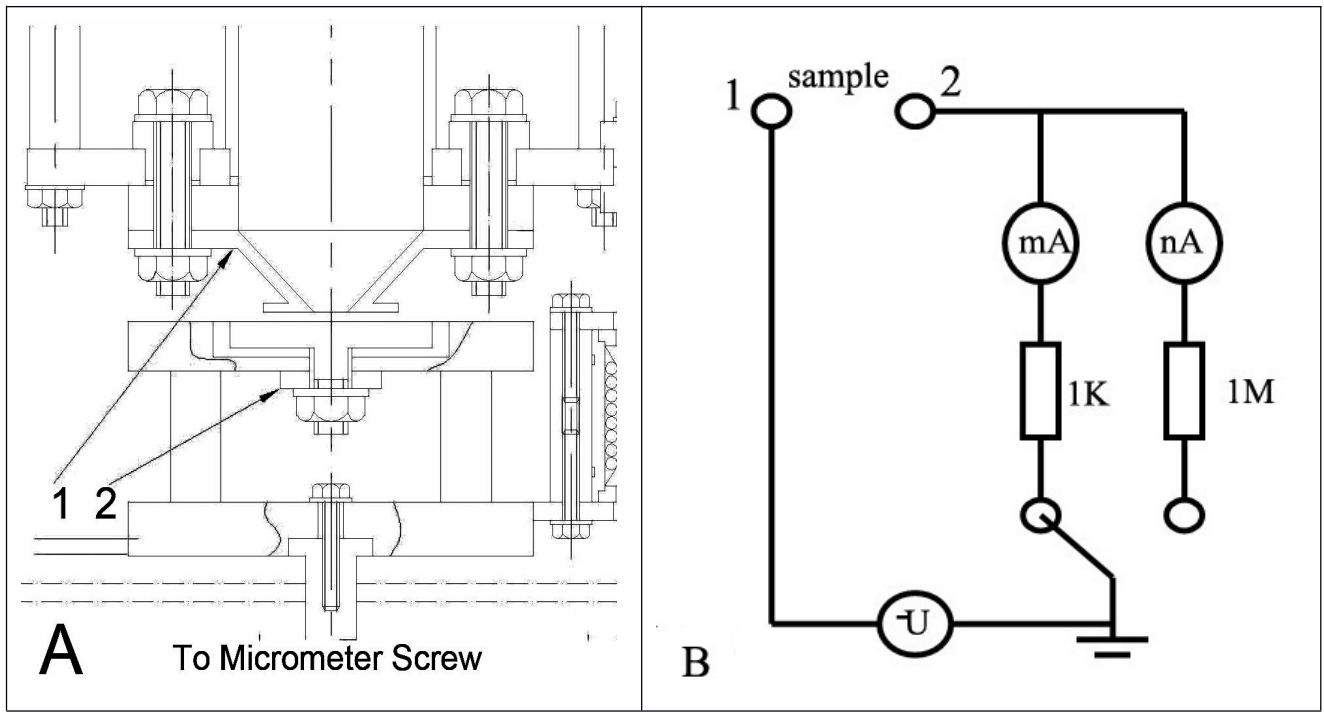

Fig 2

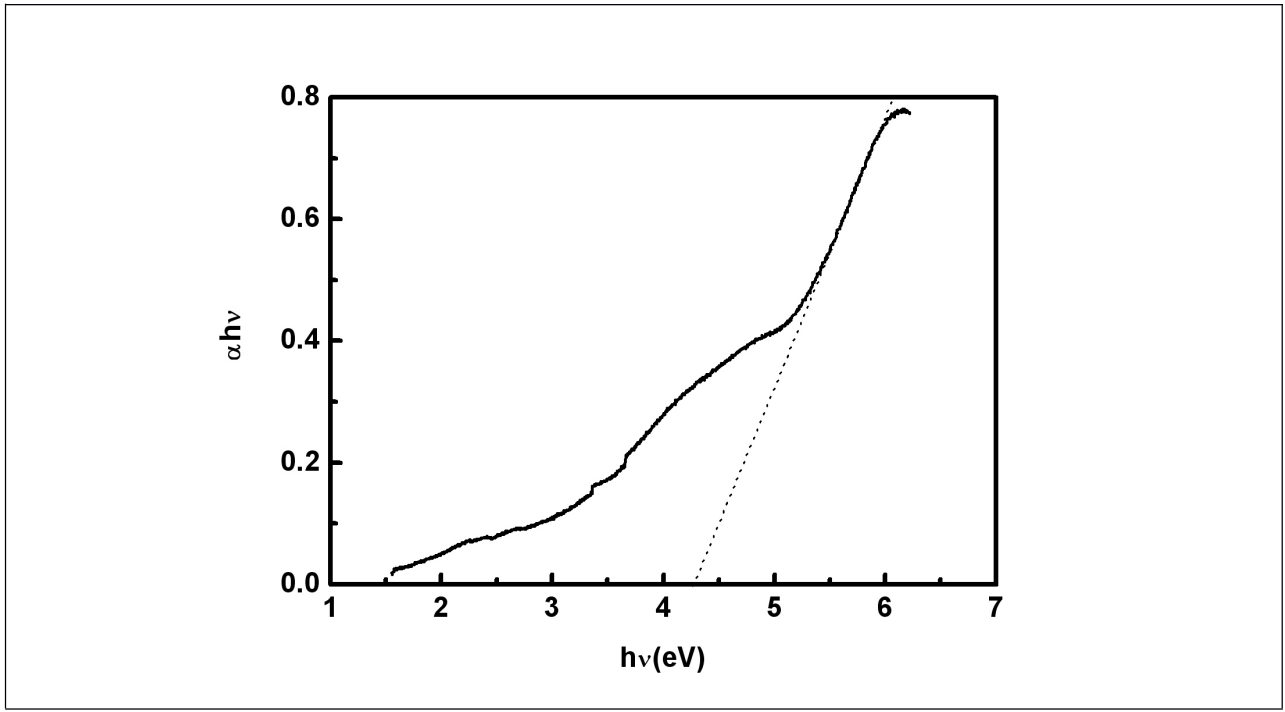

Fig 3

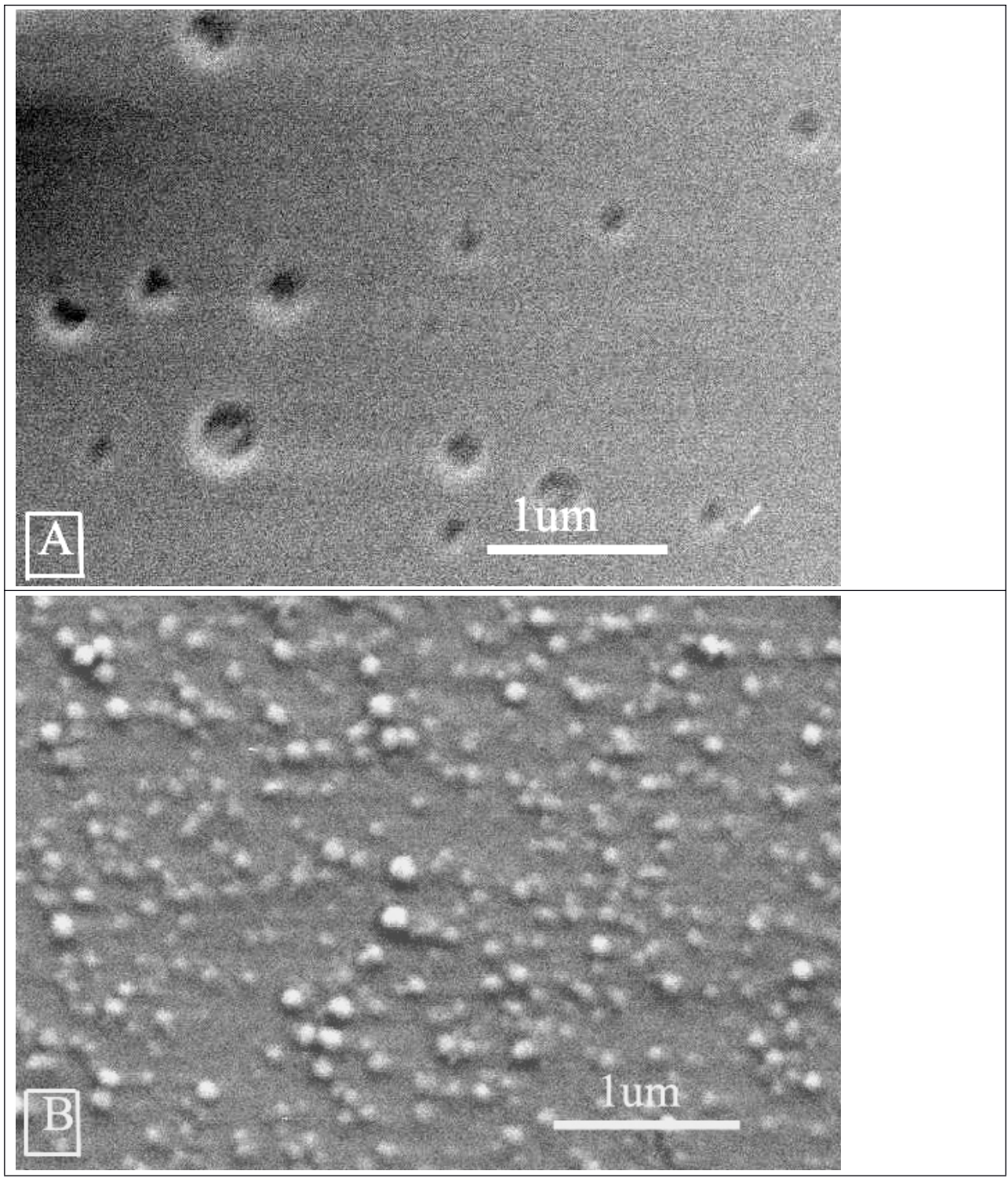

Fig 4

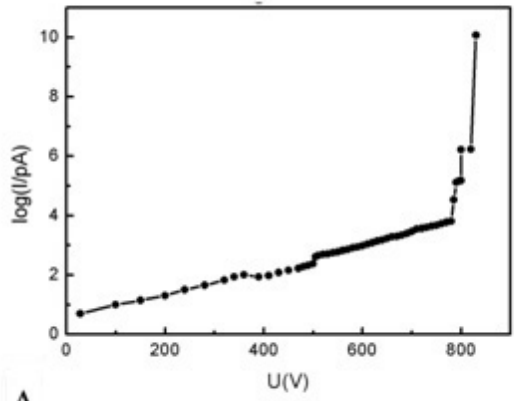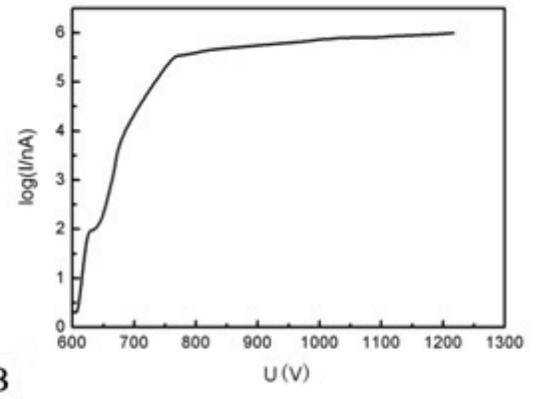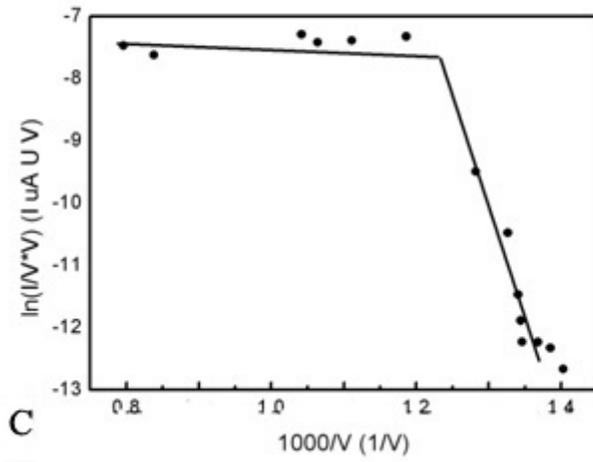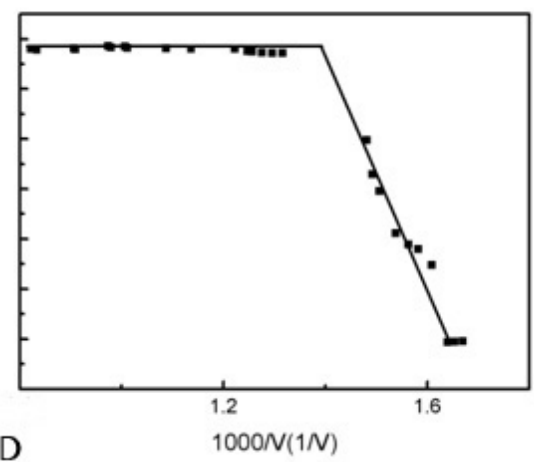